\documentclass[11pt]{article}
\usepackage{fullpage, dsfont, amsthm, amssymb, amsmath, graphicx, authblk}
\usepackage[breaklinks=false]{hyperref}

\newcommand{\captionfonts}{\small}

\makeatletter  
\long\def\@makecaption#1#2{%
  \vskip\abovecaptionskip
  \sbox\@tempboxa{{\captionfonts #1: #2}}%
  \ifdim \wd\@tempboxa >\hsize
    {\captionfonts #1: #2\par}
  \else
    \hbox to\hsize{\hfil\box\@tempboxa\hfil}%
  \fi
  \vskip\belowcaptionskip}
\makeatother   

\begin{document}

\title{Adiabatic optimization without local minima}
\author[1]{Michael Jarret}
\author[2]{Stephen P. Jordan}
\affil[1]{\small{Department of Physics, University of Maryland,
    College Park, MD 20742-4111}}
\affil[2]{\small{Applied and Computational Mathematics Division, National
  Institute of Standards and Technology, Gaithersburg, MD 20899}}

\date{}

\bibliographystyle{plain}
\maketitle

\newcommand{\braket}[2]{\langle #1|#2\rangle}
\newcommand{\bra}[1]{\langle #1|}
\newcommand{\ket}[1]{|#1\rangle}
\newcommand{\Bra}[1]{\left<#1\right|}
\newcommand{\Ket}[1]{\left|#1\right>}
\newcommand{\Braket}[2]{\left< #1 \right| #2 \right>}
\renewcommand{\th}{^\mathrm{th}}
\newcommand{\tr}{\mathrm{Tr}}
\newcommand{\id}{\mathds{1}}

\newtheorem{proposition}{Proposition}
\newtheorem{definition}{Definition}
\newtheorem{conjecture}{Conjecture}

\begin{abstract}
Several previous works have investigated the circumstances under which
quantum adiabatic optimization algorithms can tunnel out of local
energy minima that trap simulated annealing or other classical local
search algorithms. Here we investigate the even more basic question of
whether adiabatic optimization algorithms always succeed in polynomial
time for trivial optimization problems in which there are no local
energy minima other than the global minimum. Surprisingly, we find a
counterexample in which the potential is a single basin on a graph,
but the eigenvalue gap is exponentially small as a function of the
number of vertices. In this counterexample, the ground state
wavefunction consists of two ``lobes'' separated by a region of
exponentially small amplitude. Conversely, we prove if the ground
state wavefunction is single-peaked then the eigenvalue gap scales at
worst as one over the square of the number of vertices.
\end{abstract}

\section{Introduction}

In adiabatic quantum algorithms one starts with an initial Hamiltonian
whose ground state is easy to construct, such as a tensor product
state. One prepares the system in its ground state and then 
slowly varies the Hamiltonian to reach one whose ground state encodes
the solution to a computational problem of interest
\cite{Farhi_science}. The adiabatic theorem states that if the 
time-variation of the Hamiltonian is performed sufficiently slowly
then the system will track the instantaneous ground state, thereby
solving the computational problem. Specifically, for constant-rate
interpolation between the initial and final Hamiltonians, it suffices
to choose the duration of the adiabatic process to scale as
$1/\gamma^3$, where $\gamma$ is the minimal eigenvalue gap between the
ground state and first excited state during the adiabatic process
\cite{JRS07}. (More recently, it has been shown that, by instead
choosing the Hamiltonian's interpolation schedule to be a smooth
function with zero initial and final time-derivative, one can
provably achieve runtime of $\widetilde{O}(1/\gamma^2)$
\cite{Elgart_Hagedorn}.) Adiabatic quantum computation with
sufficiently general Hamiltonians can perform universal quantum
computation \cite{ADKLLR07}. However, the most natural application for
adiabatic quantum algorithms is optimization, and most analysis has
focused on this case.

One of the original intuitions behind adiabatic quantum computation
(and an earlier classical algorithm called quantum annealing
\cite{FGSSD94}) was that quantum optimization algorithms could in some
cases tunnel out of local minima that simulated annealing would fail
to climb out of. The runtime of adiabatic algorithms for various
specific potentials with local minima has been analyzed in \cite{R04,
  DMV01, FGG02, VDV03, aminchoi, Amin}. Here we investigate the more basic
question of whether quantum adiabatic algorithms always succeed in
efficiently solving ``trivial'' optimization problems that have no
local energy minima other than the global minimum. Surprisingly, we
find a counterexample in which the potential has no local minima other
than the global minimum, yet the eigenvalue gap is exponentially
small.

Specifically, we consider Hamiltonians associated with graphs,
consisting of the graph Laplacian plus a potential on the
vertices. (The dimension of the Hilbert space is the number of
vertices in the graph. The vertices may be labeled with bit strings
corresponding to basis states of a set of
qubits. Physically, one can interpret the Hamiltonian
as describing a single particle hopping amongst the vertices.) In
\S \ref{counter} we construct a single-basin potential on a graph
such that the eigenvalue gap between the ground state and first
excited state is exponentially small as a function of the number of
vertices. This corresponds to a trivial optimization problem for which
classical gradient descent finds the minimal-energy vertex in linear
time.

Strictly speaking, the exponentially small eigenvalue gap in our
example does not necessarily imply that an adiabatic algorithm fails
to solve this problem. For this one would need to invoke a converse of
the adiabatic theorem, and one would furthermore need to show that
diabatic transitions between eigenstates cause algorithmic failure in
a practical sense. (Indeed, an example of algorithmic success despite
failure of adiabaticity is given in \cite{NSK12}.) However, our
construction serves as a counterexample to a natural and perhaps even
widely assumed conjecture, namely that potentials without local
minima yield polynomial eigenvalue gaps. 

Our counterexample has a ground state consisting of two ``lobes'' with
exponentially small amplitude in the region between them. In \S
\ref{sec:conductance} we use arguments based on conductance of Markov
chains to show that the eigenvalue gap shrinks at worst quadratically
with the number of vertices provided the ground state wavefunction is
single-peaked. (See proposition \ref{mainprop}.) Thus, the two-lobed
nature of the ground state in our counterexample is an essential
feature. In other words, we find that the structure of local extrema
in the potential does not neatly characterize the eigenvalue gap, but
the structure of the local extrema of the ground state wavefunction
does.

We also specifically investigate the one-dimensional case, called the
path graph. We show that for convex\footnote{Actually, our result
  holds under the slightly weaker condition that the potential be
  ``single-basin''. See \S \ref{conductance1D}.} potentials, the
ground state wavefunction is single-peaked. This yields, as a consequence
of proposition \ref{mainprop}, an $\Omega(1/(|W|\ell^2))$ lower bound
on the gap for the path of $\ell$ vertices and a potential of norm
$|W|$. By adapting Poincar\'e's inequality we are able to obtain an
$\Omega(1/\ell^2)$ lower bound, with no dependence on $|W|$. This
lower bound is tight to within a constant factor \cite{Jarret_Jordan},
and forms a discrete analog of \cite{PW}. Previous work has shown that
for symmetric potentials on the path graph that increase as one moves
away from the center, the eigenvalue gap is lower bounded by
$\Omega(1/\ell^2)$ \cite{AB90}. Our result is incomparable to that of
\cite{AB90} in that such potentials are not a special case of convex
potentials nor vice-versa.

Much of the research on adiabatic quantum algorithms seeks to achieve
exponential speedups over classical algorithms. For this purpose, one
seeks to find a potential on a highly-connected graph of exponentially
many vertices (often the hypercube) such that the eigenvalue gap is
only polynomially small. This differs somewhat from the setting
studied in the present paper - we consider graphs of polynomially many
vertices and ask whether the gap is exponentially small or
polynomially small. Thus, our counterexample in which the gap is
already exponentially small on a graph of only polynomially many
vertices constitutes an even more extreme gap collapse than previous
examples such as \cite{AKR10}. On the other hand, our gap lower bounds
(``positive results'') might appear weak - they provide
$\Omega(1/|V_G|^2)$ lower bounds on the eigenvalue gap where $|V_G|$ is
the number of vertices in the underlying graph. However, in some
highly symmetric cases, such as optimization problems on the hypercube
with potentials that depend only on Hamming distance from the energy
minimum, the eigenvalue gap can be analyzed by collapsing the
Hamiltonian to a spectrally-equivalent Hamiltonian on an exponentially
smaller graph. (In the hypercube case, the vertices of the collapsed
graph correspond to the allowed Hamming weights, see \emph{e.g.}
\cite{Jarret_Jordan}.) Application of the tools presented here for
lower-bounding gaps in such cases remains for future work.

\section{Preliminaries}
\label{sec:preliminaries}

Let $G$ be a graph with vertices $V_G$ and edges $E_G \subseteq V_G
\times V_G$. Let $\mathcal{H}_G = \mathrm{span} \{ \ket{x} | x \in V_G \}$ be
a complex Hilbert space with $\braket{x}{y} = \delta_{x,y}$. Let $L_G$
denote the Laplacian of $G$ acting on $\mathcal{H}$. That is,
\begin{equation}
L_G = \sum_{x \in V_G} d_x \ket{x} \bra{x} - \sum_{(x,y) \in E_G} \ket{x} \bra{y},
\end{equation}
where $d_x$ denotes the degree of vertex $x$. 

The subject of spectral graph theory is devoted to analysis of the
eigenvalue spectra of graph Laplacians \cite{Chung}. Here,
motivated by applications to adiabatic quantum computation
\cite{Farhi_science}, we develop some theorems about the spectra of
more general graph-related Hamiltonians of the form
\begin{equation}
\label{HGW}
H_{G,W} = L_G + \sum_{x \in V_G} W(x) \ket{x} \bra{x},
\end{equation}
where $W:V_G \to \mathbb{R}$ is a potential energy
function. 

We say that $x \in V_G$ is a local minimum of $W$ if $W(x) \leq W(y)$ for
all $y$ such that $(x,y) \in E_G$. By the Perron-Frobenius theorem,
the ground state of $H_{G,W}$ can be expressed in the form 
\begin{equation}
\ket{\psi} = \sum_{x \in V_G} \psi(x) \ket{x}
\end{equation}
with $\psi(x) > 0$ for all $x \in V_G$. We say that $\psi$ has a local
maximum at $x$ if 
\begin{equation}
\label{psimax}
\psi(x) \geq \psi(y) \  \forall y \textrm{ s.t. } (x,y) \in E_G.
\end{equation}

In \S \ref{sec:general} we prove a lower bound on the eigenvalue
gap in the case that the ground state wavefunction is
single-peaked. By this, we mean that the set of local maxima of $\psi$ 
form a connected set of vertices in $G$. This is a weaker condition
than demanding that $\psi$ have only a single local maximum, in that
we allow the peak to consist of multiple vertices on which $\psi$ is
constant.

Most adiabatic optimization algorithms proposed to date use the
following formulation. The optimization problem is formalized as a
search on a graph $G$. The edges of the graph $E_G$ represent the
allowed moves within the search space. The vertices $V_G$ represent
the possible solutions, and one seeks to minimize the cost function
$W:V_G \to \mathbb{R}$. For simplicity we assume that $W$ has a
unique global minimum $x_{\mathrm{min}} \in V_G$. Let 
\begin{equation}
\label{HGWS}
H_{G,W}(s) = (1-s) L_G + s\sum_{x \in V_G} W(x) \ket{x} \bra{x}.
\end{equation}
The computation starts in the uniform superposition over vertices of
$G$, which is the ground state of $H_{G,W}(0)$. Then, one applies a
slowly-varying Hamiltonian $H_{G,W}(t/\tau)$. According to the
adiabatic theorem, if $\tau$ is taken sufficiently large, the system
will track the instantaneous ground state, and at the end of the
computation, one will be left with the ground state of $H(1)$, namely
$\ket{x_\mathrm{min}}$. More quantitatively, the adiabatic theorem
\cite{JRS07} shows that it suffices to take $\tau =
O(1/\gamma^3)$, where $\gamma = \min_{0 \leq s \leq 1} \gamma(s)$ and
$\gamma(s)$ is the eigenvalue gap between the ground energy and first
excited energy of $H_{G,W}(s)$. (Heuristic arguments suggest that in
many cases $\tau = O(1/\gamma^2)$ suffices \cite{Messiah}. For
careful choices of $s(t)$, which do not include the choice $s=t/\tau$
considered here, this has been shown to hold rigorously
\cite{Elgart_Hagedorn}. See \S \ref{aqc} for more discussion of
this point.)

Let
\begin{equation}
\hat{H}_{G,W}(s) = \frac{H_{G,W}(s)}{1-s}.
\end{equation}
One sees that $\hat{H}_{G,W}(s)$ is of the form (\ref{HGW}) for all $s
\in [0,1)$. Furthermore, the eigenvalue gap $\gamma(s)$ is given by
\begin{equation}
\gamma(s) = (1-s) \hat{\gamma}(s),
\end{equation}
where $\hat{\gamma}(s)$ is the eigenvalue gap of
$\hat{H}_{G,W}(s)$. Thus, theorems yielding upper or lower bounds on
the eigenvalue gap of Hamiltonians of the form (\ref{HGW}) yield useful
bounds on the eigenvalue gap of $H_{G,W}(s)$ throughout the adiabatic
algorithm except when $s$ is very close to one. The gap
analysis for $s$ very close to one can be performed by other means, as
discussed in \S \ref{aqc}. Throughout the rest of this paper, our
focus will be on bounding gaps for Hamiltonians of the form (\ref{HGW}).

Some works, such as \cite{diffinit, diffmid1, diffmid2}, have
considered adiabatic optimization algorithms with paths other than the
linear interpolation defined by (\ref{HGWS}). In certain cases this
has been shown to improve runtime. Most of the proposed alternative
paths involve non-uniform changes to the off-diagonal matrix
elements. Unlike (\ref{HGWS}), such Hamiltonians cannot be put into
the form (\ref{HGW}) by rescaling. Instead, they correspond to
(\ref{HGW}) where the Laplacian is of a weighted graph. The analysis
of such Hamiltonians thus goes beyond the scope of this paper,
although techniques related to those described here may be
applicable.

\section{Small Gaps Without Local Minima}
\label{counter}

Given a connected graph $G$, a potential $W$ on the vertices, and a
Hamiltonian $H_{G,W}$ of the form given in (\ref{HGW}), one is tempted to
conjecture that if $G$ has only polynomially many vertices and $W$ has
no local minima (other than a global minimum) then $H_{G,W}$ can't
have an exponentially small gap. In this section we construct a
counterexample to this conjecture. In fact, beyond lack of local
minima, our counterexample satisfies the even stronger condition that
the potential forms a monotonic basin leading to a unique vertex of
minimal potential. That is, there is no connected region of constant
potential.

\begin{figure}
\begin{center}
\includegraphics[width=0.65\textwidth]{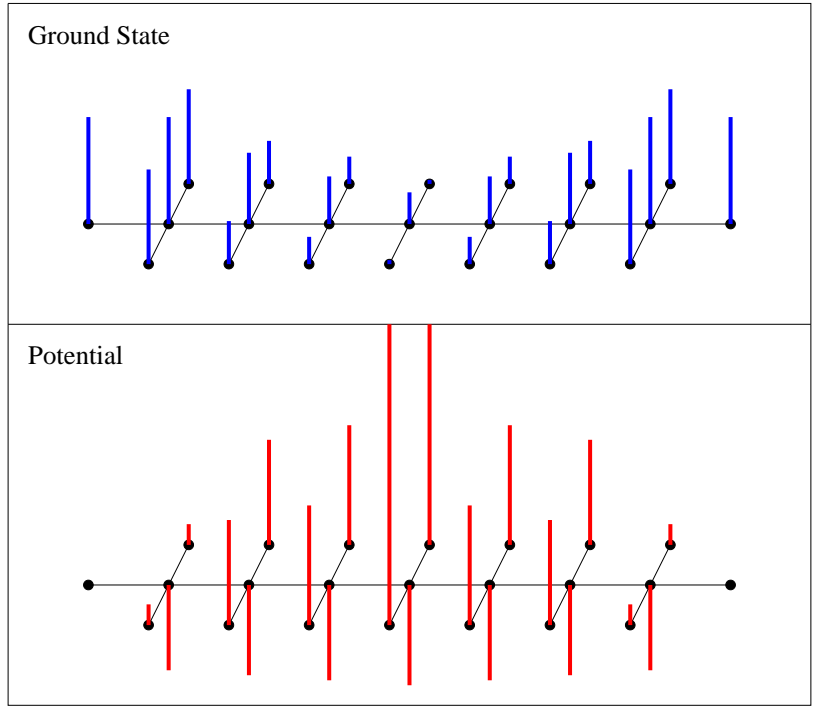}
\caption{\label{twopanes} We illustrate the ground state wavefunction
  $\psi$ and the 
  potential $W$ for $\ell = 4$. The ground state $\psi$ consists
  of two lobes separated by a region of small amplitude in the
  center. The potential along the ``spine'' of the caterpillar is
  negative and decreasing as one approaches the central vertex
  $B_4$. The  potential is positive on the ``legs'' of the
  caterpillar. Thus, the classical steepest-descent algorithm starting
  from any initial vertex will reach the minimum ($B_4$) by the
  shortest path. Note that the potential on the $C_4$ vertices is
  approximately ten times as large as the second largest value of the
  potential, and thus it is cut off by the boundaries of the figure.}
\end{center}
\end{figure}

Consider the following ``caterpillar'' graph of $6 \ell - 1$ vertices, as
illustrated below.
\[
\includegraphics[width=0.4\textwidth]{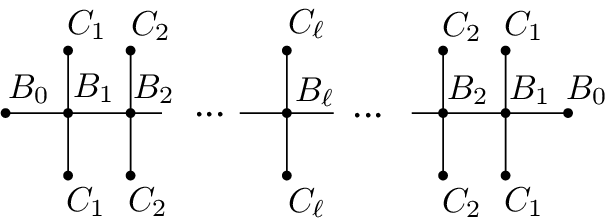}
\]
We consider a potential on the vertices with left-right and top-bottom
mirror symmetries, and we correspondingly label equivalent vertices with
identical labels. Our potential is as follows\footnote{Curious readers
  may wonder how this potential was arrived at. One can choose a
  desired ground state and potential on the $B$ vertices, set the
  ground energy to zero, and solve for the wavefunction and potential
  on the $C$ vertices. With some trial and error one can find
  choices such that the wavefunction at $B_\ell$ is exponentially
  small, yet the potential on each $C$ vertex is greater than the
  potential on the $B$ vertex to which it is connected and the ground
  state amplitudes are nonnegative on all vertices.}.
\begin{equation}
\begin{array}{rcll}
W(B_0) & = & 0 & \vspace{5pt} \\
W(B_j) & = & -\frac{1}{2} - \frac{j}{4l} & j \in \{1,\ldots,\ell\}
\vspace{5pt} \\
W(C_1) & = & \frac{1}{\frac{11}{12}-\frac{1}{8 \ell}}-1 & \vspace{5pt}\\
W(C_\ell) & = & 7 & \vspace{5pt} \\
W(C_j) & = & \frac{1}{\frac{2}{3}-\frac{j}{8 \ell}} -1 & j \in
\{2,\ldots,\ell-1\}
\end{array}
\end{equation}
One sees that this potential is a single basin funneling to the unique
minimum-potential vertex $B_\ell$. (See Fig. \ref{twopanes}.) The
following unnormalized eigenstate has eigenvalue zero.
\begin{equation}
\label{psi}
\begin{array}{rcll}
\psi(B_0) & = & \frac{2}{3} & \vspace{5pt} \\
\psi(B_j) & = & \left( \frac{2}{3} \right)^j & j \in \{1,\ldots,\ell\}
\vspace{5pt} \\
\psi(C_\ell) & = & \frac{1}{8} \left( \frac{2}{3} \right)^\ell &
\vspace{5pt} \\
\psi(C_1) & = & \frac{2}{3} \left( \frac{11}{12}-\frac{1}{8 \ell}
\right)  & \vspace{5pt} \\
\psi(C_j) & = & \left( \frac{2}{3}-\frac{j}{8 \ell} \right) \left(
  \frac{2}{3} \right)^j & j \in \{2,\ldots,\ell-1\}
\end{array}
\end{equation}
All off-diagonal elements of the Hamiltonian $H_{G,W}$ are
nonpositive. Therefore, by the Perron-Frobenius theorem, its ground
state is the only eigenstate with all nonnegative amplitudes
\cite{BDOT08}. Hence, we can identify $\psi$ as the ground state of
$H_{G,W}$.

A ground state consisting of two symmetric lobes, such as $\psi$,
implies a small eigenvalue gap because, by flipping the signs of the
amplitudes in one lobe, one obtains an orthogonal state of only
slightly higher energy. This energy cost, which upper-bounds the
eigenvalue gap, is small due to the smallness of the amplitudes
between the lobes.

More precisely, consider the wavefunction $\phi$, which equals $\psi$ for all
vertices to the left of $B_{\ell}$, equals $-\psi$ for all vertices
to the right of $B_{\ell}$, and equals zero at $B_{\ell}$ and
$C_\ell$. One sees that $\phi$ is orthogonal to $\psi$. Let $\eta =
\braket{\phi}{\phi}$ and let $\ket{\widetilde{\phi}} =
\frac{1}{\sqrt{\eta}} \ket{\phi}$ be the normalized version of
$\ket{\phi}$. The first excited state is variationally characterized
as the lowest energy state orthogonal to the ground state. Therefore
the energy of the first excited state is at most
$\bra{\widetilde{\phi}} H_{G,W} \ket{\widetilde{\phi}}$. Because the
ground energy is zero we thus have
\begin{equation}
\label{gammaphi}
\gamma \leq \bra{\widetilde{\phi}} H_{G,W} \ket{\widetilde{\phi}}.
\end{equation}

By construction, $\ket{\phi}$ satisfies the 
eigenvalue zero equation everywhere except at the $B_{\ell}$ vertex
and the two $B_{\ell-1}$ vertices. Using this fact, one finds
\begin{equation}
\bra{\phi} H_{G,W} \ket{\phi} = 2 \psi(B_{\ell}) \psi(B_{\ell-1}).
\end{equation}
By (\ref{psi}) one sees that $\eta > 1$. Therefore,
(\ref{gammaphi}) yields 
\begin{eqnarray}
\gamma & \leq & 2 \eta^{-1} \psi(B_{\ell})\psi(B_{\ell-1}) \\
       & < & 2 \psi(B_{\ell}) \psi(B_{\ell-1}) \\
       & = & 2 \left( \frac{2}{3} \right)^{2\ell-1}.
\end{eqnarray}
Hence, without any local minima in the potential and with only
$O(\ell)$ vertices we obtain an eigenvalue gap of $O((2/3)^{2\ell})$. 

\section{Conductance-based Gap Bounds}
\label{sec:conductance}

In the preceding section, we showed that a ground state consisting of
two symmetric lobes separated by a region of small amplitude implies a
small eigenvalue gap. We relied on the symmetry of the lobes to
construct a low-energy state orthogonal to the ground state by
flipping the sign of the amplitudes on one lobe. However, it is true
more generally that lobes separated by a region of small amplitude
imply a small gap even if the lobes are asymmetric, provided the
imbalance is not too severe. In this section we use concept of
conductance to make this precise, and conversely to prove that if the
ground state wavefunction is single-peaked, then the eigenvalue gap
cannot be smaller than $\Omega(|V_G|^{-2})$.

\subsection{Conductance}
\label{subsec:conductance}

Motivated by applications to rapidly mixing Markov chains,
sophisticated tools have been developed to bound the difference
between the largest and second-largest eigenvalues of stochastic 
matrices. In this subsection, we recount one such tool, known as
conductance.

Consider a discrete-time random walk on $G$ defined by transition
matrix $P$. That is, for $x,y \in V_G$, $P_{xy}$ is the
probability for a walker at $x$ to transition to $y$ in
a given timestep. Thus, $P$ is a row-stochastic matrix. Conductance
provides upper and lower bounds on the gap between the largest and
second largest eigenvalues of row-stochastic matrices in the case that
the random walks they define are ergodic and reversible. Ergodicity means
that the random walk converges to the same limiting distribution
independent of the starting point of the walker. Reversibility means
that, in the limiting distribution, the probability of traversing a
given edge in one direction is equal to the probability of traversing
it in the opposite direction. More formally, we recount the following
definitions and facts from \cite{Sinclair}.

\begin{definition}
The random walk defined by transition matrix $P$ on vertex set $V_G$
is ergodic if
\begin{equation}
\lim_{s \to \infty} \left( P^s \right)_{xy} = \pi_y \quad
\textrm{independent of $x$}.
\end{equation}
The probability distribution $\pi$ is then called the limiting
distribution of the random walk.
\end{definition}

\begin{proposition}
\label{ergodicity}
The following conditions are necessary and sufficient for ergodicity
of $P$.
\begin{enumerate}
\item $P$ is irreducible. That is, for each $x,y \in V_G$ there is $s
  \in \mathbb{N}$ such that $\left( P^s \right)_{xy} > 0$.
\item $P$ is aperiodic. That is, for all $x,y$, $\mathrm{gcd} \{ s|
  (P^s)_{xy} > 0 \} = 1$.
\end{enumerate}
\end{proposition}

\begin{definition}
An ergodic random walk given by transition matrix $P$ on vertex set
$V_G$ is reversible if
\begin{equation}
\pi_x P_{xy} =  \pi_y P_{yx} \quad \forall x,y \in V_G,
\end{equation}
where $\pi$ is the limiting distribution.
\end{definition}

\begin{definition}
Let $P$ be the transition matrix of a reversible ergodic random walk
on graph $G$ with vertices $V_G$ and edges $E_G$. Let $\pi$ be the
corresponding limiting distribution. Let $S$ be any non-empty subset
of $V_G$ and let $\bar{S} = V_G/S$ be its complement. Let
\begin{eqnarray}
F_S & = & \mathop{\sum_{(x,y) \in E_G}}_{x \in S, y \in \bar{S}} \pi_x
P_{xy} \vspace{30pt} \\[5pt]
C_S & = & \sum_{x \in S} \pi_x \\[5pt]
\Phi_S(P) & = & \frac{F_S}{\min\{C_S, C_{\bar{S}}\}} \\[5pt]
\Phi(P) & = & \min_{S \subset V_G} \Phi_S(P).
\end{eqnarray}
$\Phi(P)$ is called the conductance of $P$.
\end{definition}

The quantity $F_S$ is called the flow of $S$, and the quantity $P_S$
is called the probability of $S$. Note that, for reversible random
walks, $F_S = F_{\bar{S}}$. By the Perron-Frobenius theorem, the
largest eigenvalue of any irreducible stochastic matrix is 1 and the
corresponding eigenspace is one-dimensional. Furthermore, this
eigenvector can be written with all nonnegative entries. Adapting
theorems 2.4 and 2.6 of \cite{Sinclair} one has the following.

\begin{proposition}
\label{conductance}(from \cite{Sinclair})
Let matrix $P$ define a reversible ergodic random walk with
conductance $\Phi(P)$. Let $\gamma$ denote the gap between the largest
eigenvalue of $P$ (which is 1) and the second-largest eigenvalue. Then
\begin{equation}
\frac{\Phi(P)^2}{2} \leq \gamma \leq 2 \Phi(P).
\end{equation}
\end{proposition}

Proposition \ref{conductance} is based on Cheeger's inequality
\cite{Cheeger} for the spectrum of Laplacians of manifolds, which was
adapted to graphs by Alon and Milman \cite{Alon}, and extended to
stochastic matrices by Sinclair \cite{Sinclair}.

\subsection{Conductance Bound}
\label{sec:general}

In this subsection we use conductance to prove lower bounds on the gap
of Hamiltonians of the form $H_{G,W}$ given in (\ref{HGW}),
culminating in a proof that the ``lobed'' nature of the ground state
wavefunction in the counterexample from \S \ref{counter} is a
necessary feature to obtain exponentially small gap. Specifically, we
show that if $H_{G,W}$ has a single-peaked ground state then its
eigenvalue gap has an $\Omega(|W|^{-1} |V_G|^{-2})$ lower bound, where
$|V_G|$ is the number of vertices in the graph $G$ and $|W| =
\max_{x\in V_G} W(x) - \min_{x \in V_G} W(x)$.

Given a connected graph $G$, and a potential $W$ on the vertices, let
$H_{G,W}$ be the corresponding Hamiltonian of the form (\ref{HGW}).
Let $\gamma$ denote the energy gap between the ground state and first
excited state of $H_{G,W}$. For the purpose of bounding $\gamma$ we
may assume without loss of generality that the potential satisfies
$W(x) < -d_G \quad \forall x \in V_G$, where $d_G$ is the maximum degree
of any vertex in $G$. If this is not the case, one can always
subtract a sufficiently large multiple of the identity matrix to make
it so without affecting $\gamma$.

Let $\ket{\psi} = \sum_{x \in V_G} \psi(x) \ket{x}$ denote the ground
state of $H_{G,W}$ and $E$ the ground energy. Let $N_x$ be the
neighbors of vertex $x$. That is, 
\begin{equation}
N_x = \{ y \in V_G| (x,y) \in E_G\}.
\end{equation}
In this notation,
\begin{equation}
\label{eigen1}
 (d_x +W(x)) \psi(x) - \sum_{y \in N_x} \psi(y) = E \psi(x).
\end{equation}
For connected $G$, 
\begin{equation}
\label{strictly}
\psi(x) > 0 \quad \forall x.
\end{equation}
Thus we may rearrange
(\ref{eigen1}) to obtain
\begin{equation}
\label{eigen2}
d_x + W(x) -\sum_{y \in N_x} \psi(y)/\psi(x)  = E.
\end{equation}
Also, note that $H$ has all nonpositive entries, so $E < 0$. 

We next adapt a technique from \cite{ATS03, BT09, AP10} to relate the
spectrum of $H_{G,W}$ to the spectrum of a random walk. Let $D =
\mathrm{diag}\{\psi(x)|x \in V_G\}$. By (\ref{strictly}), $D$ is an
invertible matrix with $D^{-1} = \mathrm{diag}\{\psi(x)^{-1}|x \in
V_G\}$. Let
\begin{equation}
\label{Pdef}
P = \frac{1}{E} D^{-1} H_{G,W} D.
\end{equation}
By (\ref{eigen2}), $\sum_{y \in V_G} \bra{x} P \ket{y} = 1$. That is,
$P$ is a row-stochastic matrix.

Because $E < 0$, the lowest eigenvalue of $H$ corresponds to the
highest eigenvalue of $P$, which is 1. Specifically, let
\begin{equation}
\ket{\psi^2} = \sum_{x \in V_G} \psi(x)^2 \ket{x}.
\end{equation}
One sees that
\begin{equation}
\bra{\psi^2} P = \bra{\psi^2}.
\end{equation}
Hence the probability distribution $\psi^2$ is a limiting distribution
of the random walk defined by $P$. Connectedness of the graph $G$
suffices to ensure that condition 1 of proposition \ref{ergodicity} is
satisfied. The requirement that $W(x) < -d_G$ for all $x
\in V_G$ ensures that condition 2 of proposition \ref{ergodicity} is
satisfied \cite{Sinclair}. Thus, $P$ is an ergodic random walk. In
other words, $\psi^2$ is the unique limiting distribution of $P$ and
correspondingly $\ket{\psi}$ is the nondegenerate ground state of
$H_{G,W}$. By direct calculation, one finds
\begin{equation}
\psi(x)^2 P_{xy} = \psi(y)^2 P_{yx} = \left \{ \begin{array}{cl}
-\frac{1}{E} \psi(x) \psi(y) & \textrm{if $(x,y) \in E_g$}\\
0 & \textrm{otherwise}.
\end{array} \right.
\end{equation}
Thus, $P$ is a reversible ergodic random walk. Therefore, by
proposition \ref{conductance} and equation (\ref{Pdef}), the energy
gap $\gamma$ between the ground and first-excited states of $H_{G,W}$
satisfies
\begin{equation}
\label{bounds}
-\frac{E}{2} \Phi^2(P) \leq \gamma \leq -2E \Phi(P).
\end{equation}
One sees that the flow between $S \subset V_G$ and its complement
determined by $P$ is
\begin{equation}
F_S(P) = \mathop{\sum_{x \in S}}_{y \in \bar{S}} \frac{\psi(x)
  \psi(y)}{-E}
\end{equation}
and the corresponding probability is
\begin{equation}
C_S(P) = \sum_{x \in S} \psi(x)^2.
\end{equation}
Thus, by (\ref{bounds}) one obtains the following result.
\begin{proposition}
\label{mainlemma}(cf. \cite{ATS03, BT09, AP10})
Let $H_{G,W}$ be a Hamiltonian of the form (\ref{HGW}) with $W(x) \leq
-d_G \quad \forall x \in V_G$. Let $\psi$ denote the ground state of
$H_{G,W}$, let $E$ denote the ground energy, and let $\gamma$ denote
the gap between the ground energy and the first excited energy. Then,
\begin{equation}
\label{lowerbound}
 - \frac{1}{2E} \Phi_H^2 \leq \gamma \leq 2 \Phi_H
\end{equation}
where
\begin{eqnarray}
\Phi_H & = & \min_{S \subset V_G} \frac{F_S}{\min
    \{C_S, C_{\bar{S}}\} } \\
F_S & = & \sum_{(x,y) \in B} \psi(x) \psi(y) \label{ar1}\\
B   & = & \{(x,y)| x \in S, y \notin S, (x,y) \in E_G\} \label{ar2}\\
C_S & = & \sum_{x \in S} \psi(x)^2 \label{ar3}\\
C_{\bar{S}} & = & \mathop{\sum_{x \in V_G}}_{x \notin S} \psi(x)^2. \label{ar4}
\end{eqnarray}
\end{proposition}
\noindent
Note that $E < 0$ and therefore the lower bound on $\gamma$ given by
(\ref{lowerbound}) is nonnegative.

Examining (\ref{lowerbound}) one sees that the gap is exponentially
small if and only if the ground state has a pair of 
not-too-unbalanced lobes separated by a region of exponentially small
amplitude. Choosing $S$ and $\bar{S}$ to be the lobes, one sees that
$S$ and $\bar{S}$ must have reasonably well-balanced ground state
probabilities for the denominator $\min\{C_S, C_{\bar{S}}\}$ to remain
large, and the amplitudes along the cut separating $S$ from $\bar{S}$
must all be small for the numerator $F_S$ to be small. More precisely,
recalling from \S \ref{sec:preliminaries} the definition of
single-peaked, we have the following, which is the main result of this
section.

\begin{proposition}
\label{mainprop}
Let $G$ be a connected graph with vertices $V_G$, edges $E_G$, and
maximum degree $d_G$. Let $W:V_G \to \mathbb{R}$ be a potential, and
$H_{G,W}$ the corresponding Hamiltonian described in (\ref{HGW}). Let
$\psi$ denote the ground state of $H_{G,W}$ and let $\gamma$ denote
the eigenvalue gap between the ground state and first excited state of
$H_{G,W}$. If $\psi$ is single-peaked then
\begin{equation}
\gamma \geq \frac{1}{2 (|W|+d_G) |V_G|^2}
\end{equation}
where
\begin{equation}
|W| = \max_{x \in V_G} W(x) - \min_{x \in V_G} W(x).
\end{equation}
\end{proposition}

\noindent
\begin{proof}
Let 
\begin{equation}
H_{G,W}^{(-)} = H_{G,W} - (W_{\max}+d_G) \id
\end{equation} 
where $W_{\max} = \max_{x \in V_G} W(x)$. One sees that
$H_{G,W}^{(-)}$ has the same ground state $\psi$ and same gap $\gamma$
as $H_{G,W}$ and that all matrix elements in $H_{G,W}^{(-)}$ are
nonpositive. Hence, by 
proposition \ref{mainlemma},
\begin{equation}
\label{first}
\gamma \geq - \frac{1}{2E^{(-)}} \left( \min_{S \subset V_G} \frac{F_S}{\min\{C_S,
    C_{\bar{S}}\}} \right)^2
\end{equation}
where $E^{(-)}$ is the ground energy of $H_{G,W}^{(-)}$, namely
\begin{equation}
E^{(-)} = E - (W_{\max}+d_G),
\end{equation}
and $F_S$, $C_S$, and $C_{\bar{S}}$ are as in
(\ref{ar1})-(\ref{ar4}). Graph Laplacians are positive semidefinite,
and therefore $E \geq W_{\min}$. Thus,
\begin{equation}
E^{(-)} \geq -|W|-d_G.
\end{equation}
Hence, (\ref{first}) yields
\begin{equation}
\label{second}
\gamma \geq \frac{1}{2(|W|+d_G)} \left( \min_{S \subset V_G}
  \frac{F_S}{\min\{C_S, C_{\bar{S}}\}} \right)^2.
\end{equation}
We now consider two cases: 1) the peak of $\psi$ spans the
cut $\{S,\bar{S}\}$, and 2) the peak of $\psi$ is contained entirely
within one side of the cut.\\
\textbf{Case 1:}
If the peak of $\psi$ spans the cut then there exist $x \in S$ and $y
\in \bar{S}$ such that $(x,y) \in E_G$ and $\psi(x) = \psi(y) \geq
\psi(z) \ \forall z \in V_G$. We can lower bound $\gamma$ by throwing
away the flows across all edges in the numerator other than
$(x,y)$. Thus,
\begin{equation}
\gamma \geq \frac{1}{2(|W|+d_G)} \left( \frac{\psi(x)^2}
{\min\{C_{S},C_{\bar{S}}\}} \right)^2.
\end{equation}
Furthermore, $\min\{C_{S},C_{\bar{S}}\} \leq \psi(x)^2 |V_G|$,
and therefore $\gamma \geq \frac{1}{2 (|W|+d) |V_G|^2}$.\\
\textbf{Case 2:}
If the peak of $\psi$ is contained within one side of the cut, we may,
without loss of generality, call the side containing the peak $S$ and
the other side $\bar{S}$. Let $x_{\max}$ be the vertex in $\bar{S}$
that maximizes $\psi$. Because $\psi$ is single-peaked, there must be
a neighbor $z$ of $x_{\max}$ such that $\psi(z) >
\psi(x_{\max})$. Because $\psi(x_{\max})$ maximizes $\psi$ in
$\bar{S}$, $z$ must be contained in $S$. We can lower bound $\gamma$
by throwing away the flows across all edges in the numerator other
than $(x_{\max},z)$. Thus,
\begin{equation}
\gamma \geq \frac{1}{2(|W|+d_G)} \left( \frac{\psi(x_{\max}) \psi(z)}
{\min\{C_{S},C_{\bar{S}}\}} \right)^2
\geq \frac{1}{2(|W|+d_G)} \left( \frac{\psi(x_{\max})^2}
{\min\{C_{S},C_{\bar{S}}\}} \right)^2.
\end{equation}
Furthermore, $C_{\bar{S}} \leq \psi(x_{\max})^2 |V_G|$, and therefore
$\min\{C_{S},C_{\bar{S}}\} \leq \psi(x_{\max})^2 |V_G|$. 
Thus, in this case also, $\gamma \geq \frac{1}{2 (|W|+d_G) |V_G|^2}$.
\end{proof}

\subsection{Conductance Bound for Path Graphs}
\label{conductance1D}

Here we note some consequences of proposition \ref{mainprop} in the
case that $G$ is the path graph of $l$ vertices, $G_l$.
\[
\begin{array}{rcl}
G_l & = & \includegraphics[width=0.2\textwidth]{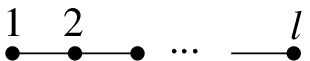}
\end{array} \vspace{10pt}
\]

\begin{definition}
Let $G$ be a graph with vertices $V_G$ and edges $E_G$. Let $W:V_G \to
\mathbb{R}$ be a potential. We say $W$ is a single-basin potential
if the set $\{x \in V_G| W(x) < E\}$ is a connected set of vertices in
$G$ for all $E$.
\end{definition}

As we now show, single-basin potentials on the path graph have
single-peaked ground states and hence a large eigenvalue gap by
proposition \ref{mainprop}. For intuition, recall that, for a single
particle in the one-dimensional continuum, the time-independent
Schr\"odinger equation can be written as 
$-\frac{d^2 \psi}{dx^2} = (E-W(x)) \psi$. The ground state can be
expressed with all real non-negative amplitudes. Hence the sign of
$\frac{d^2 \psi}{dx^2}$ is the same as the sign of $W(x)-E$. Thus, the
ground state of a convex potential has simple structure: inside the
well, $W(x)-E < 0$ and the wavefunction is concave down, whereas
outside the well $W(x)-E > 0$ and the wavefunction is concave up. The
path graph case, described below, is essentially a discrete analogue
to this. \\

\noindent
\textbf{Remark:} The notion of a single-basin potential is
well-defined on any graph. On path graphs one can also easily
define the notion of a convex potential. Simply think of the $l$
vertices as corresponding to the integers $\{1,\ldots,l\}$ and demand
that the potential on the vertices be equal to some convex function on
$\mathbb{R}$ evaluated at these integer points. It is not hard to show
that single-basin is a slightly weaker condition than convex. That is,
on the path graph, all convex potentials are single-basin, but not all
single-basin potentials are convex. \\

For a wavefunction $\psi$ on the vertices of $G$, define
\begin{equation}
\Delta^2 \psi(x) = - d_x \psi(x) + \sum_{y \in N_x} \psi(y),
\end{equation}
where $d_x$ is the degree of vertex $x$ and $N_x$ is the set vertices
neighboring $x$. Thus, 
\begin{equation}
\label{deriv}
L_G \ket{\psi} = -\sum_{x \in V_G} \Delta^2 \psi(x) \ket{x}.
\end{equation}

\begin{proposition}
Suppose $W$ is a single-basin potential on graph $G$. Let $\psi$ be
the ground state of the corresponding Hamiltonian $H_{G,W}$, and let
\begin{equation}
S[\psi] = \{x \in V_G| \Delta^2 \psi(x) < 0 \}.
\end{equation}
Then, $S[\psi]$ is a connected set of vertices in $G$.
\end{proposition}

\noindent
\begin{proof}
Let $E$ denote the ground energy of $H_{G,W}$. Thus, by (\ref{deriv}),
\begin{equation}
\Delta^2 \psi(x) = (W(x)- E) \psi(x)
\end{equation}
Recall that $\psi(x) > 0 \quad \forall x \in V_G$. Thus, $\Delta^2
\psi(x)$ has the same sign as $W(x)- E$. The connectedness of
$S[\psi]$ then follows directly from the single-basin property.
\end{proof}

In special case that $G$ is a path graph, the connectedness
of $S[\psi]$ implies that $\psi$ has only one local
maximum. Thus, as a corollary of proposition \ref{mainprop}, one
obtains proposition \ref{1Dprop}. Note that on more general graphs,
connectedness of $S[\psi]$ does not imply that $\psi$ has only one
local maximum.

\begin{proposition}
\label{1Dprop}
Let $W$ be a single-basin potential on the path graph $G_l$. Let
$H_{G,W}$ be the corresponding Hamiltonian of the form
(\ref{HGW}). Let $\gamma$ denote the gap between the ground energy and
first excited energy of $H_{G,W}$. Then $\gamma \geq
\frac{1}{2(|W|+2)l^2}$ where $|W| = \max_{x \in V_G} W(x) - \min_{x
  \in V_G} W(x)$.
\end{proposition}

Proposition \ref{1Dprop} shows that for single-basin potentials on
$G_l$, the eigenvalue gap obeys $\gamma = \Omega(1/l^2)$. In the
special case of a flat potential, it is easy to solve for the
eigenvalue gap exactly, which is $O(1/l^2)$. However, the bound of
proposition \ref{1Dprop} is not tight due to the dependence on
$|W|$. In the next section, we obtain a tighter bound by applying
the Poincar\'e inequality.

\section{Poincar\'e-based Gap Bounds}

Two of the main tools for proving lower bounds on the eigenvalue gap
of stochastic matrices are the Cheeger inequality and the Poincar\'e
inequality. Conductance methods, such as those described in \S
\ref{subsec:conductance}, are originally derived from the Cheeger
inequality \cite{Cheeger}. For some random walks, the Poincar\'e
inequality yields stronger lower bounds than the Cheeger inequality
\cite{DS91, FW99}, and for other random walks the reverse is true
\cite{P12}. In \S \ref{sec:poincare}, we recount the version of
the Poincar\'e inequality given in \cite{DS91} and apply it to
Hamiltonians $H_{G,W}$ of the form (\ref{HGW}). In \S
\ref{sec:poinpath} we specialize to the case of path graphs, obtaining
a tighter bound than our conductance-based bound (proposition
\ref{1Dprop}). (For a previous example in which Poincar\'e's
inequality is used to bound the gap of a Hamiltonian see
\cite{BCMNS12}.)

\subsection{The Poincar\'e Inequality}
\label{sec:poincare}

Let $P$ be the transition matrix for an ergodic reversible
discrete-time random walk on a graph $G$. Let $\pi$ denote the
limiting distribution and let $\gamma$ denote the gap between the
highest and second-highest eigenvalues of $P$. For
any edge $e$ in the graph $G$, let $e_1,e_2$ denote the vertices at
its endpoints. Let $Q(e)$ denote the flow across edge $e$ in the
limiting distribution.
\begin{equation}
Q(e) = \pi_{e_1} P_{e_1,e_2} = \pi_{e_2} P_{e_2,e_1}.
\end{equation} 
The latter equality expresses the reversibility of the random walk.
For each ordered pair $(x,y)$ of distinct vertices in $G$, choose a
canonical path $\gamma_{xy}$ from $x$ to $y$. Vertices may be repeated
in a path, but no edge may be traversed more than once. Let $\Gamma$
be the collection of canonical paths, one for each ordered pair of
vertices. For
$\gamma_{xy} \in \Gamma$, let
\begin{equation}
|\gamma_{xy}| = \sum_{e \in \gamma_{xy}} Q(e)^{-1}
\end{equation}
where the sum is over the edges in path $\gamma_{xy}$. Let
\begin{equation}
\kappa(\Gamma) = \max_e \sum_{\gamma_{xy} \owns e} |\gamma_{xy}|
\pi_x \pi_y.
\end{equation}
The Poincar\'e inequality states \cite{DS91}
\begin{equation}
\label{poincare}
\gamma \geq \frac{1}{\kappa}.
\end{equation}
To obtain a tight bound on $\gamma$ one must make a good choice of
$\Gamma$.

Intuitively, the quantity $\frac{1}{\kappa}$, like the conductance
$\Phi$, quantifies the presence of a bottleneck across which the flow
is small. As an example, consider a graph consisting of two large
subgraphs connected by only a single edge $e$. In this case, every pair of
vertices spanning the pair of subgraphs has a canonical path crossing
$e$. Correspondingly, $\sum_{\gamma_{xy} \owns e} |\gamma_{xy}| \pi_x \pi_y$ will
be large, which implies large $\kappa$. Similarly, $\kappa$ will be
large if there are many edges connecting the two subgraphs to each
other but the flow $Q(e)$ across all such edges is small. Only in the
absence of such bottlenecks does (\ref{poincare}) yield a large lower
bound on the gap.

As in \S \ref{sec:general}, we use (\ref{Pdef}) to obtain a
stochastic matrix $P$ from our Hamiltonian $H$ such that the
eigenvalue gap $\gamma$ of $P$ relates to the eigenvalue gap
$\gamma_H$ of $H$ according to 
\begin{equation}
\label{gammarel2}
\gamma_H = -E \gamma,
\end{equation}
where $E$ is the ground energy of $H$. The eigenvalue gap of
$P$ can be lower-bounded using the Poincar\'e
inequality. Specifically, by (\ref{Pdef}), we have the following.
\begin{eqnarray}
Q(x,y) & = & \frac{\psi(x)\psi(y)}{-E} \\
\pi_x & = & \psi(x)^2 \\
\kappa & = & \max_e \sum_{\gamma_{xy} \owns e} \psi(x)^2 \psi(y)^2
\sum_{g \in \gamma_{xy}} \frac{-E}{\psi(g_1) \psi(g_2)}.
\end{eqnarray}
Here $\psi$ is the ground state of $H$, and $g_1,g_2$ are the two
vertices connected by edge $g$. By (\ref{gammarel2}) the ground energy
cancels from the final bound on  $\gamma_H$. Summarizing:
\begin{equation}
\label{gammahpoin}
\gamma_H \geq \frac{1}{\kappa'},
\end{equation}
where
\begin{equation}
\label{kappaprime}
\kappa' = \max_e \sum_{\gamma_{xy} \owns e} \psi(x)^2 \psi(y)^2
\sum_{g \in \gamma_{xy}} \frac{1}{\psi(g_1) \psi(g_2)}.
\end{equation}

\subsection{Poincar{\'e} Bound for Path Graphs}
\label{sec:poinpath}

For path graphs, there is only one valid choice of canonical paths
$\Gamma$. Specifically, for a pair of vertices $s < f$ the canonical
path is $s,s+1,\ldots,f$. For $f < s$ one takes the reverse
path. Thus, (\ref{kappaprime}) reduces to 
\begin{equation}
\label{simprime}
\kappa' = \max_{1 \leq j \leq l-1} 2 \sum_{s \leq j} \sum_{f > j}
R(s,f)
\end{equation}
where
\begin{equation}
\label{RSF}
R(s,f) = \psi(s)^2 \psi(f)^2 \sum_{s \leq v < f} \frac{1}{\psi(v) \psi(v+1)}.
\end{equation}
The factor of 2 in (\ref{simprime}) arises because we sum only over
the paths with $s < f$ and use the fact that $R(s,f) = R(f,s)$.
 
As discussed in \S \ref{conductance1D}, if the potential on the
path graph is single-basin, then the ground state wavefunction has only one
local maximum. Thus, the minimum of $\psi(v)$ along a segment $s \leq
v < f$ must occur at one of the endpoints. If the minimum is at $s$
then (\ref{RSF}) yields
\begin{eqnarray}
R(s,f) & \leq & \psi(s)^2 \psi(f)^2 \sum_{s \leq v < f}
\frac{1}{\psi(s)^2} \\
& = & (f-s) \psi(f)^2.
\end{eqnarray}
Similarly, if the minimum is at $f$ then one has $R(s,f) \leq (f-s)
\psi(s)^2$.

Let $J$ be the value of $j$ that achieves the maximum in
(\ref{simprime}). Then
\begin{equation}
\kappa' \leq 2 \sum_{s \leq J} \sum_{f > J}  (f-s) \psi(b_{s,f})^2
\end{equation}
where $b_{s,f}$ is either $s$ or $f$ depending on which is smaller amongst
$\psi(s)^2$ and $\psi(f)^2$. We can rewrite this sum over pairs of
vertices as
\begin{equation}
\label{rejigger}
\sum_{s \leq J} \sum_{f > J}  (f-s) \psi(b_{s,f})^2 = \sum_{b=1}^l
\sum_{a \in S_b} |a-b| \psi(b)^2,
\end{equation}
where, for a given vertex $b$, $S_b$ is the set of
vertices on the other side of edge $J$ such that $\psi(a)^2 \leq
\psi(b)^2$. (For some $b$, $S_b$ can be empty.) From (\ref{rejigger})
we have
\begin{eqnarray}
\kappa' & \leq & 2 \sum_{b=1}^l \psi(b)^2 \sum_{a \in S_b} |a-b| \\
        & \leq & 2 \sum_{b=1}^l \psi(b)^2 \sum_{a=1}^{l-1} a \\
        & =    & \sum_{b=1}^l \psi(b)^2 l(l-1) \\
        & \leq & l (l-1).
\end{eqnarray}
The last equality follows from the fact that $\psi(b)^2$ is a
probability distribution over $1,\ldots,l$. Thus, by
(\ref{gammahpoin}),
\begin{equation}
\label{final}
\gamma_H \geq \frac{1}{l(l-1)}.
\end{equation}
By direct calculation, one finds that the eigenvalue gap for the
length $l$ chain with no potential ($W=0$) is $4 \sin^2 \left(
  \frac{\pi}{2 l} \right)$. Thus, the bound (\ref{final}) is
asymptotically tight to within a factor of $\pi^2$
\cite{Jarret_Jordan}.

\section{Application to Adiabatic Optimization Algorithms}
\label{aqc}

In this section, we show that, as a corollary of proposition
\ref{mainprop}, adiabatic optimization algorithms in which the ground
state $\psi(s)$ is single-peaked for all $s$, have minimum gap at
least $\Omega(1/|V_G|^2)$ and therefore run in
$\widetilde{O}(|V_G|^4)$ time, by an adiabatic theorem
\cite{Elgart_Hagedorn}. (The $\widetilde{O}$ notation indicates that we
are omitting logarithmic factors.) This result cannot be used directly
to find algorithmic speedups, as exhaustive search runs in $O(|V_G|)$
time. However, we believe this analysis may be useful in cases of high
symmetry such as \cite{R04, DMV01, FGG02}, where the eigenvalue gap
on exponentially large graphs can be determined by analyzing the spectrum
of polynomial-size graphs. In addition, the analysis in this section
provides an illustrative example of how proposition \ref{mainprop} may
be applied to the analysis of adiabatic optimization problems.

Consider an adiabatic optimization algorithm using a Hamiltonian
$H_{G,W}(s)$ of the form shown in (\ref{HGWS}). Then 
\begin{equation}
\hat{H}_{G,W}(s) = \frac{1}{1-s} H_{G,W}(s)
\end{equation}
is of the form (\ref{HGW}) addressed by proposition
\ref{mainprop}. $\hat{H}_{G,W}(s)$ and $H_{G,W}(s)$ have the same
ground state, which we denote $\psi(s)$. Thus, if $\psi(s)$ is
single-peaked for all $s \in [0,1)$ we may conclude from proposition
\ref{mainprop} that
\begin{equation}
\label{raw_conclusion}
\hat{\gamma}(s) \geq \frac{1}{2 \left( |\hat{W}(s)|+d_G \right) |V_G|^2},
\end{equation}
where $\hat{W}(s) = \frac{s}{1-s}W$ is the potential in
$\hat{H}_{G,W}(s)$. Hence, one substitutes $|\hat{W}(s)| =
\frac{s}{1-s} |W|$ and $\gamma(s) = (1-s) \hat{\gamma}(s)$ into
(\ref{raw_conclusion}), obtaining 
\begin{equation}
\label{gammabound}
\gamma(s) \geq \frac{1-s}{2 \left( \frac{s}{1-s} |W| + d_G \right)
  |V_G|^2}.
\end{equation} 

One sees that this lower bound on $\gamma(s)$ becomes very small as
$s$ closely approaches 1. For the final part of the adiabatic
optimization algorithm we therefore use a different method to
lower-bound the eigenvalue gap. As an illustrative example, we suppose
that the gap between the minimum of $W$ and the second smallest value
taken by $W$ is one. Thus, by (\ref{HGWS}), $\gamma(1) =
1$. Generalization to other values of $\gamma(1)$ is straightforward and
yields the same scaling with $|V_G|$ and $d_G$. At $s = 1 - \delta$,
one has
\begin{equation}
H(s) = \delta L_G + (1-\delta) W.
\end{equation}
By Gershgorin's circle theorem, one sees that the operator norm of
$L_G$ is at most $2 d_G$. Thus, the operator norm of $\delta L_G$ is
at most $2 \delta d_G$. Hence, Weyl's inequalities show that the worst
case is that the addition of $\delta L_G$ to $(1-\delta)W$ shifts the
ground energy up by $2 \delta d_G$ and shifts the first excited energy
down by $2 \delta d_G$. Thus, adding $\delta L_G$ to $(1-\delta) W$ at
worst decreases the gap from $1-\delta$ to $1-\delta-4 \delta
d_G$. Thus,
\begin{equation}
\gamma(s) \geq \frac{1}{2}-\frac{1}{8 d_G} \quad \forall s \in
\left[1-\frac{1}{8 d_G},1\right].
\end{equation}
The degree $d_G$ is at least 2 for any connected graph of more than
two vertices, so for all nontrivial cases one has
\begin{equation}
\label{gammaend}
\gamma(s) \geq \frac{7}{16} \quad \forall s \in
\left[1-\frac{1}{8 d_G},1\right].
\end{equation}
For the remaining values of $s$, (\ref{gammabound}) yields
\begin{equation}
\label{gammastart}
\gamma(s) \geq \frac{\frac{1}{8 d_G}}
{2 \left( 8 d_G |W| + d_G \right) |V_G|^2} 
\quad \forall s \in \left[0,1-\frac{1}{8 d_G} \right].
\end{equation}
Together, (\ref{gammaend}) and (\ref{gammastart}) yield
\begin{equation}
\label{allgamma}
\gamma(s) = \Omega \left( \frac{1}{d_G^2 |W| |V_G|^2} \right)
\quad \forall s \in [0,1].
\end{equation}
The adiabatic theorem of \cite{JRS07} shows that adiabaticity will be
maintained by evolving according to the linear-interpolation
Hamiltonian $H_{G,W}(t/\tau)$ with runtime $\tau$ bounded by
\begin{equation}
\label{runtime}
\tau = O \left( \frac{ \left\| \frac{dH}{ds} \right\|^2}{\gamma^3} \right).
\end{equation}
By (\ref{HGWS}), $\left\| \frac{dH}{ds} \right\| = O(d_G + |W|)$. Thus,
by (\ref{allgamma}) and (\ref{runtime}),
\begin{equation}
\tau = O \left( d_G^6 |W|^3 |V_G|^6 (|W|+d_G)^2 \right).
\end{equation}

As shown in \cite{Elgart_Hagedorn}, a tighter bound on running time
can be obtained by choosing a more optimized interpolation schedule
between the initial and final Hamiltonians. Specifically, one should
choose the interpolation such that $H(t)$ is infinitely differentiable
but is time-independent outside of $t \in [0,\tau]$. For example,
let
\begin{equation}
H(t) = (1-s(t/\tau)) L_G + s(t/\tau) W
\end{equation}
where $s$ is the following ``switching function'', which is infinitely
differentiable, and satisfies $s(0) = 0$, $s(1) = 1$, and
$s'(x) = 0 \  \forall x \notin (0,1)$:
\begin{eqnarray}
s(x) & = & \int_{-\infty}^x g(y) dy \\
g(y) & = & \left\{ \begin{array}{ll}
0 & \textrm{if $y \notin [0,1]$} \\
\beta \exp \left( - \frac{1}{y(1-y)} \right) & \textrm{if $y \in (0,1)$}
\end{array} \right. . \\
\end{eqnarray}
Here, $\beta$ is the normalization constant yielding $f(1) = 1$.
In this case, as shown in \cite{Elgart_Hagedorn}, by evolving with
$H(t)$ from time zero to $\tau$ one achieves adiabaticity with
runtime
\begin{equation}
\label{switching_adiabatic}
\tau = O\left( \frac{(\log(1/\gamma))^{12}}{\gamma^2} \right).
\end{equation}
For a Hamiltonian in which the ground state is always single-peaked,
(\ref{switching_adiabatic}) and (\ref{allgamma}) yield runtime
\begin{equation}
\label{fast}
\tau = \widetilde{O}(V_G^4).
\end{equation}

\section{Concluding Remarks}

The examples analyzed here and in \cite{R04, DMV01, FGG02, VDV03,
  aminchoi} show that quantum adiabatic algorithms can succeed in
finding the minimum in polynomial time in cases where classical local
search fails to do so, and it can fail in cases where classical local
search succeeds. For both classical local search and adiabatic
optimization, local minima of the potential that one is seeking to
minimize play an important role in determining runtime. However, as
the present work shows, these local minima do not tell the whole
story. In particular, absence of local minima does not imply large
eigenvalue gap.

In addition, we note that there remains much to be learned regarding
the performance of adiabatic optimization algorithms relative to
classical computation in the general case that one is not comparing
only to classical local search. In particular, the classical algorithm
described in appendix A of \cite{childswalk} finds the minimum in
polynomial time for most of the known examples in which adiabatic
optimization beats classical local search. We hope that the tools
developed here will be helpful in investigating this issue.\\ 

\noindent
\textbf{Acknowledgments: }We thank Amanda Streib, Noah Strieb, and
Alexey Gorshkov for useful conversations. Portions of this paper are a
contribution of NIST, an agency of the US government, and are not
subject to US copyright. This work was supported in part by the center
for Quantum Information and Computer Science (QuICS).

\bibliography{optimization}

\begin{thebibliography}{10}

\bibitem{ATS03}
Dorit Aharonov and Amnon Ta-Shma.
\newblock Adiabatic quantum state generation and statistical zero knowledge.
\newblock In {\em STOC '03: Proceedings of the thirty-fifth annual ACM
  Symposium on Theory of Computing}, pages 20--29, 2003.
\newblock arXiv:quant-ph/0301023.

\bibitem{ADKLLR07}
Dorit Aharonov, Wim van Dam, Julia Kempe, Zeph Landau, Seth Lloyd, and Oded
  Regev.
\newblock Adiabatic quantum computation is equivalent to standard quantum
  computation.
\newblock {\em SIAM Journal on Computing}, 37(1):166--194, 2007.
\newblock arXiv:quant-ph/0405098.

\bibitem{AP10}
Abbas Al-Shimary and Jiannis~K. Pachos.
\newblock Energy gaps of {H}amiltonians from graph {L}aplacians.
\newblock {\em arXiv:1010.4130}, 2010.

\bibitem{Alon}
N.~Alon and V.~D. Milman.
\newblock {$\lambda_1$}, isoperimetric inequalities for graphs, and
  superconcentrators.
\newblock {\em Journal of Combinatorial Theory, Series B}, 38:73--88, 1985.

\bibitem{AKR10}
Boris Altshuler, Hari Krovi, and J{\'e}r{\'e}mie Roland.
\newblock Anderson localization makes adiabatic quantum optimization fail.
\newblock {\em Proceedings of the National Academy of Sciences},
  107(28):12446--12450, 2010.

\bibitem{Amin}
M.~H.~S. Amin.
\newblock Effect of local minima on adiabatic quantum optimization.
\newblock {\em Physical Review Letters}, 100:130503, 2008.
\newblock arXiv:0709.0528.

\bibitem{aminchoi}
M.~H.~S. Amin and V.~Choi.
\newblock First order quantum phase transition in adiabatic quantum
  computation.
\newblock {\em Phyisical Review A}, 80:062326, 2009.
\newblock arXiv:0904.1387.

\bibitem{AB90}
Mark~S. Ashbaugh and Rafael~D. Benguria.
\newblock Some eigenvalue inequalities for a class of {J}acobi matrices.
\newblock {\em Linear Algebra and its Applications}, 136:215--234, 1990.

\bibitem{BCMNS12}
Sergey Bravyi, Libor Caha, Ramis Movassagh, Daniel Nagaj, and Peter Shor.
\newblock Criticality without frustration for quantum spin-1 chains.
\newblock {\em Physical Review Letters}, 109:207202, 2012.
\newblock arXiv:1203.5801.

\bibitem{BDOT08}
Sergey Bravyi, David~P. DiVincenzo, Roberto Oliveira, and Barbara~M. Terhal.
\newblock The complexity of stoquastic local {H}amiltonian problems.
\newblock {\em Quantum Information and Computation}, 8(5):361--385, 2008.
\newblock arXiv:quant-ph/0606140.

\bibitem{BT09}
Sergey Bravyi and Barbara Terhal.
\newblock Complexity of stoquastic frustration-free {H}amiltonians.
\newblock {\em SIAM Journal on Computing}, 39(4):1462, 2009.
\newblock arXiv:0806.1746.

\bibitem{Cheeger}
Jeff Cheeger.
\newblock A lower bound for the smallest value of the {L}aplacian.
\newblock In {\em Problems in Analysis}, pages 195--199. Princeton University
  Press, 1970.

\bibitem{childswalk}
Andrew~M. Childs, Richard Cleve, Enrico Deotto, Edward Farhi, Sam Gutmann, and
  Daniel~A. Spielman.
\newblock Exponential algorithmic speedup by a quantum walk.
\newblock In {\em STOC '03: Proceedings of the thirty-fifth annual ACM
  Symposium on Theory of Computing}, pages 59--68, 2003.
\newblock arXiv:quant-ph/0209131.

\bibitem{Chung}
Fan R.~K. Chung.
\newblock {\em Spectral Graph Theory}.
\newblock Number~92 in Regional Conference Series in Mathematics. American
  Mathematical Society, 1997.

\bibitem{diffmid2}
Elizabeth Crosson, Edward Farhi, Cedric Yen-Yu Lin, Han-Hsuan Lin, and Peter
  Shor.
\newblock Different strategies for optimization using the quantum adiabatic
  algorithm.
\newblock {\em arXiv:1401.7320}, 2014.

\bibitem{DS91}
Persi Diaconis and Daniel Stroock.
\newblock Geometric bounds for eigenvalues of {M}arkov chains.
\newblock {\em The Annals of Applied Probability}, 1(1):36--61, 1991.

\bibitem{Elgart_Hagedorn}
Alexander Elgart and George~A. Hagedorn.
\newblock A note on the switching adiabatic theorem.
\newblock {\em Journal of Mathematical Physics}, 53:102202, 2012.

\bibitem{diffinit}
Edward Farhi, Jeffrey Goldstone, David Gosset, Sam Gutmann, Harvey~B. Meyer,
  and Peter Shor.
\newblock Quantum adiabatic algorithms, small gaps, and different paths.
\newblock {\em Quantum Information and Computation}, 11:181--214, 2011.
\newblock arXiv:0909.4766.

\bibitem{FGG02}
Edward Farhi, Jeffrey Goldstone, and Sam Gutmann.
\newblock Quantum adiabatic evolution algorithms versus simulated annealing.
\newblock {\em arXiv:quant-ph/0201031}, 2002.

\bibitem{diffmid1}
Edward Farhi, Jeffrey Goldstone, and Sam Gutmann.
\newblock Quantum adiabatic evolution algorithms with different paths.
\newblock {\em arXiv:quant-ph/0208135}, 2002.

\bibitem{Farhi_science}
Edward Farhi, Jeffrey Goldstone, Sam Gutmann, Joshua Lapan, Andrew Lundgren,
  and Daniel Preda.
\newblock A quantum adiabatic evolution algorithm applied to random instances
  of an {NP}-complete problem.
\newblock {\em Science}, 292(5516):472--475, 2001.
\newblock arXiv:quant-ph/0104129.

\bibitem{FGSSD94}
A.~B. Finnila, M.~A. Gomez, C.~Sebenik, C.~Stenson, and J.~D. Doll.
\newblock Quantum annealing: a new method for minimizing multidimensional
  functions.
\newblock {\em Chemical Physics Letters}, 219:343--348, 1994.

\bibitem{FW99}
J.~Fulman and E.~L. Wilmer.
\newblock Comparing eigenvalue bounds for {M}arkov chains: when does
  {Poincar\'e} beat {Cheeger}?
\newblock {\em Annals of Applied Probability}, 9(1):1--13, 1999.

\bibitem{JRS07}
Sabine Jansen, Mary-Beth Ruskai, and Ruedi Seiler.
\newblock Bounds for the adiabatic approximation with applications to quantum
  computation.
\newblock {\em Journal of Mathematical Physics}, 48:102111, 2007.
\newblock arXiv:quant-ph/0603175.

\bibitem{Jarret_Jordan}
Michael Jarret and Stephen~P. Jordan.
\newblock Fundamental gap for a class of {Schr\"odinger} operators on path and
  hypercube graphs.
\newblock {\em Journal of Mathematical Physics}, 55(5):052104, 2014.
\newblock arXiv:1403.1473.

\bibitem{Messiah}
Albert Messiah.
\newblock {\em Quantum Mechanics}.
\newblock Dover, 1961.

\bibitem{NSK12}
Daniel Nagaj, Rolando~D. Somma, and Maria Kieferova.
\newblock Quantum speedup by quantum annealing.
\newblock {\em Physical Review Letters}, 109:050501, 2012.
\newblock arXiv:1202.6257.

\bibitem{PW}
L.~E. Payne and H.~F. Weinberger.
\newblock An optimal {P}oincar{\'e} inequality for convex domains.
\newblock {\em Archive for Rational Mechanics and Analysis}, 5(1):286--292,
  1960.

\bibitem{P12}
John Pike.
\newblock A note on the {Poincar\'e and Cheeger} inequalities for simple random
  walk on a connected graph.
\newblock {\em arXiv:1210.5777}, 2012.

\bibitem{R04}
Ben Reichardt.
\newblock The quantum adiabatic optimization algorithm and local minima.
\newblock In {\em Proceedings of STOC '04}, pages 502--510, 2004.

\bibitem{Sinclair}
Alistair Sinclair.
\newblock {\em Algorithms for random generation and counting: a Markov chain
  approach}.
\newblock Birkhauser, 1993.

\bibitem{DMV01}
Wim van Dam, Michele Mosca, and Umesh Vazirani.
\newblock How powerful is adiabatic quantum computation?
\newblock In {\em Proceedings of FOCS '01}, pages 279--287, 2001.
\newblock arXiv:quant-ph/0206003.

\bibitem{VDV03}
Wim van Dam and Umesh Vazirani.
\newblock Limits of quantum adiabatic optimization.
\newblock \url{www.cs.berkeley.edu/~vazirani/pubs/qao.pdf}, 2003.

\end{thebibliography}

\end{document}